\documentclass[12pt,a4paper,oneside]{article}
\usepackage{ifpdf}
\usepackage{a4wide}
\usepackage{amsmath}
\usepackage{graphicx}
\usepackage{rotating}
\usepackage[numbers,sort&compress]{natbib} 
\usepackage{amssymb}
\begin{document}

\textheight 21.0cm
\textwidth 16cm
\sloppy
\oddsidemargin 0.0cm \evensidemargin 0.0cm
\topmargin 0.0cm

\setlength{\parskip}{0.45cm}
\setlength{\baselineskip}{0.75cm}



\begin{titlepage}
\setlength{\parskip}{0.25cm}
\setlength{\baselineskip}{0.25cm}
\begin{flushright}
DO-TH 09/13\\
\vspace{0.2cm}
August 2009
\end{flushright}
\vspace{1.0cm}
\begin{center}
\Large
{\bf Variable Flavor Number Parton Distributions and 
Weak Gauge and Higgs Boson Production\\
at Hadron Colliders at NNLO of QCD}
\vspace{1.5cm}

\large
P.~Jimenez-Delgado and E.\ Reya
\vspace{1.0cm}

\normalsize
{\it Universit\"{a}t Dortmund, Institut f\"{u}r Physik}\\
{\it D-44221 Dortmund, Germany} \\

\vspace{1.5cm}
\end{center}

\begin{abstract}
\noindent 
Based on our recent NNLO dynamical parton distributions as obtained
in the  `fixed flavor number scheme', we generate radiatively parton
distributions in the  `variable flavor number scheme' where also
the heavy quark flavors ($c,b,t$) become massless partons within the
nucleon.  Only within this latter factorization scheme NNLO
calculations are feasible at present, since the required partonic
subprocesses are only available in the approximation of massless
initial--state partons.  The NNLO predictions for gauge boson
production are typically {\em larger} (by more than $1\sigma$) than
the NLO ones, and rates at LHC energies can be predicted with an
accuracy of about 5\%, whereas at Tevatron they are more than $2\sigma$
above the NLO ones.  The NNLO predictions for SM Higgs boson production
via the dominant gluon fusion process have a total (pdf and scale)
uncertainty of about 10\% at LHC which almost doubles at the lower
Tevatron energies; they are typically about 20\% larger than the ones
at NLO but the total uncertainty bands overlap.
\end{abstract}
\end{titlepage}

\section{Introduction}
Parton distributions and their implications have been recently 
studied within the dynamical (radiative) parton model approach up
to next-to-next-to-leading order (NNLO) of QCD \cite{ref1}.  
Here the predicted steep small Bjorken-$x$
behavior of structure functions is mainly due to QCD-dynamics at 
\mbox{$x$ \raisebox{-0.1cm}{$\stackrel{<}{\sim}$} $10^{-2}$,} 
since the parton distributions at 
$Q^2$ \raisebox{-0.1cm}{$\stackrel{>}{\sim}$} 1 GeV$^2$ are
QCD radiatively generated from {\em valencelike} positive definite
input distributions at an optimally determined low input scale
$Q_0^2\equiv \mu^2<1$ GeV$^2$. 
(`Valencelike' refers to $a_f>0$ for {\em all} input distributions
$xf(x,\mu^2)\propto x^{a_f}(1-x)^{b_f}$, i.e., not only the valence
but also the sea and gluon input densities vanish at small 
$x$).\footnote{Alternatively, in the common ``standard'' approach the 
input scale is fixed at some arbitrarily chosen $Q^2_0>1$ GeV$^2$
and the corresponding input distributions are less restricted. 
For example, the observed {\em steep} small-$x$ behavior ($a_f<0$)
of structure functions and consequently of the gluon and sea
distributions has to be {\em fitted}. Furthermore the associated
uncertainties encountered in the determination of the parton 
distributions turn out to be larger, particularly in the small-$x$
region, than in the more restricted dynamical radiative approach where,
moreover, the evolution distance (starting at $Q_0^2<1$ GeV$^2$) is
sizably larger (see, e.g., \cite{ref1} and references therein.)}
Such analyses are usually performed within the framework of the 
so-called  `fixed flavor number scheme' (FFNS) where, besides the
gluon, only the light quark flavors $q=u,d,s$ are considered 
as genuine, i.e., massless partons within the nucleon.  
This factorization scheme is fully predictive in the heavy quark
$h=c,b,t$ sector where the heavy quark flavors are produced entirely
perturbatively as final state quantum fluctuations in the strong
field generated by the initial light quarks and gluons.
Here the full heavy quark mass $m_{c,b,t}$ dependence is taken into
account in the production cross sections, as required experimentally
\cite{ref2,ref3,ref4,ref5}, in particular, in the threshold region. 
However, even for very large values of $Q^2$, $Q^2\gg m_{c,b}^2$,
these FFNS predictions up to next-to-leading order (NLO) are in
remarkable agreement \cite{ref6,ref7} with deep inelastic scattering
(DIS) data and, moreover, are perturbatively stable despite the common
belief that ``noncollinear'' logarithms $\ln(Q^2/m_h^2)$ have to be
resummed for $h=c,b$, and eventually $t$.
This agreement with experiment even at $Q^2\gg m_h^2$ indicates that 
there is {\em little} need to resum these supposedly ``large logarithms'',
which is of course in contrast to the genuine collinear logarithms 
appearing in light (massless) quark and gluon hard scattering processes.
It should be mentioned that, so far, the heavy NNLO 
${\cal{O}}(\alpha_s^3)$ 3-loop corrections to $F_{2,L}$ have been 
calculated only asymptotically for $Q^2\gg m_h^2$ 
\cite{ref8,ref9,ref10,ref11}.

In many situations, calculations within this factorization scheme 
become unduly complicated (for a recent discussion, see \cite{ref12}).
Thus it is advantageous to consider the so-called ``variable flavor
number scheme'' (VFNS) despite the somewhat questionable resummations
of heavy quark mass effects using massless evolution equations, starting
at unphysical ``thresholds'' $Q^2=m_h^2$.  
Here the heavy quarks 
($c,b,t$) are considered to be massless partons within the nucleon as
well, with their distributions $h(x,Q^2)=\bar{h}(x,Q^2)$ being generated,
up to NLO, from the boundary conditions $h(x,m_h^2)=\bar{h}(x,m_h^2)=0$,
and at NNLO from $h(x,m_h^2)=\bar{h}(x,m_h^2)={\cal{O}}(\alpha_s^2)$ as
will be explicitly given in the next Section.  
Thus this factorization
scheme is characterized by increasing the number of flavors $n_f$ of
massless partons by one unit at $Q^2=m_h^2$ starting from $n_f=3$ at
$Q^2=m_c^2$.  Hence the $n_f>3$ ``heavy'' quark distributions are
perturbatively uniquely generated from the $n_f-1$ ones via the massless
renormalization group $Q^2$ evolutions (see, e.g.\ \cite{ref13,ref14};
a comparative qualitative and quantitative discussion of this zero-mass
VFNS and the FFNS has been recently presented in \cite{ref12}). 
Eventually one nevertheless has to {\em assume} that these massless
``heavy'' quark distributions are relevant asymptotically and that they
correctly describe the asymptotic behavior of DIS structure functions
for scales $Q^2\gg m_h^2$.  However, for most experimentally accessible
values of $Q^2$, in particular around the threshold region of heavy
quark $(h\bar{h})$ production, effects due to {\em finite} heavy quark
masses $m_h$ can {\em not} be neglected.  One therefore needs an improvement
of this zero-mass VFNS where heavy quark mass-dependent corrections are
maintained in the hard cross sections. Such improvements are generally
referred to as the general-mass VFNS and there exist various different
model-dependent ways of implementing the required $m_h$ dependence
\cite{ref15,ref16,ref17,ref18,ref19,ref20,ref21,ref22,ref23,ref24,ref25,
ref26,ref27,ref28}.\footnote{Notice that it is rather superfluous to 
argue about the best
choice of a factorization scheme since the scheme choice remains merely
a theoretical convention as long as there are no observable signatures
which allow to uniquely distinguish between the FFNS and any version of
a general-mass VFNS (except the strictly massless VFNS which has been
known to be experimentally inadequate for a very long time.)}
\noindent These factorization schemes interpolate between the zero-mass VFNS 
(assumed to be correct asymptotically) and the (experimentally required)
FFNS used for our previous analysis \cite{ref1}.

In order to avoid any such model ambiguities we shall generate in the next
Section the ``heavy'' zero-mass VFNS distributions using our unique
NNLO dynamical FFNS distributions \cite{ref1} as input at $Q^2=m_c^2$.
This will considerably ease the otherwise unduly complicated calculations
in the FFNS of gauge- and Higgs-boson production and heavy quark production
at collider energies, or the calculation of weak charged-current 
(anti)neutrino-nucleon cross sections at ultrahigh neutrino energies,
for example.  
It has been recently shown \cite{ref12} that for situations
where the invariant mass of the produced system ($cW,\, tW,\, t\bar{b},\,$
Higgs-bosons, etc.) exceeds by far the mass of the participating heavy
flavor, the VFNS predictions deviate rather little from the FFNS ones,
typically by about 10\% which is within the margins of renormalization and 
factorization scale uncertainties, and ambiguities related to presently
available parton distributions.  
Let us consider, for example, hadronic $W^{\pm}$ production. 
The relevant heavy quark contributions at LO have to be calculated via
$g\bar{s}(\bar{d})\to\bar{c}W^+$, $gu\to bW^+$ in the (fully massive)
FFNS as compared to the much simpler quark fusion subprocesses
$c\bar{s}(\bar{d})\to W^+$,
$\bar{b}u\to W^+$ in the VFNS, etc. 
Here nonrelativistic contributions from the threshold region in the 
FFNS are suppressed due to $\sqrt{\hat{s}_{th}}/m_{c,b}\simeq M_W/m_{c,b}
\gg 1$. Similarly, hadronic single top production via $W$-gluon fusion
\cite{ref29} requires in the FFNS the calculation of the subprocess
$ug\to dt\bar{b}$ at LO and of $ug\to dt\bar{b}g$, etc., at NLO; in the
VFNS one needs merely $ub\to dt$ at LO and $ub\to dtg$, etc., at NLO, 
using massless initial-state partons.
Again, $\sqrt{s_{th}}/m_b\simeq m_t/m_b \gg 1$ and thus the FFNS and
VFNS results are not too different.  
A similar agreement is obtained for hadronic (heavy) Higgs boson production
where the LO FFNS subprocess $gg\to b\bar{b}H$ has to be compared
with the $b\bar{b}$ fusion subprocess (for massless initial-state partons)
 in the VFNS starting with $b\bar{b}\to H$ at LO.  
($H=H_{\rm SM}^0; h^0,\, H^0,\, A^0$ denote the Standard Model (SM) 
Higgs boson or a light scalar $h^0$, a heavy scalar $H^0$ and a pseudoscalar
$A^0$ of supersymmetric theories with 
$M_H$ \raisebox{-0.1cm}{$\stackrel{>}{\sim}$} 100 GeV.)
Again, $\sqrt{\hat{s}_{th}}/m_b=(2m_b+M_H)/m_b\gg 1$ in the FFNS which 
indicates that the simpler LO, NLO, and NNLO VFNS $b\bar{b}$ fusion
processes do provide reliable predictions.
(Notice that these situations are very different from DIS heavy quark
$h\bar{h}$ production via $\gamma^* g\to h\bar{h}$, etc., where 
$\sqrt{\hat{s}_{th}}/m_h =2$ is not sufficiently large to exclude 
significant contributions from the threshold region and therefore the
VFNS predictions deviate sizably from the FFNS ones \cite{ref12}).

Within the present intrinsic theoretical uncertainties we can therefore
rely on our uniquely generated NNLO VFNS parton distribution functions
(pdfs) where, moreover, the required NNLO cross sections for massless
initial-state partons are, in contrast to the fully massive FFNS, available
in the literature for a variety of important production processes. 
The perturbative stability of the NNLO predictions, when compared with 
the ones based on our dynamical NLO VFNS pdfs \cite{ref12}, will be 
furthermore studied in the next Section for the hadronic production of
$W^{\pm}$ and $Z^0$ bosons, as well as of the SM Higgs boson at the 
Tevatron and at LHC.
Our conclusions are summarized in Sect.\ 3.  
Finally, the Mellin $n$-moments of the renormalized heavy-quark flavor
operator matrix elements relevant for the generation of the VFNS pdfs
at NNLO are summarized in the Appendix.

\section{Heavy flavor parton distributions and their implications at 
high energy colliders}

As common, the flavor transitions $n_f\to n_f+1$ are made when the 
factorization scale equals the (pole) mass of the heavy quarks, 
$Q^2=m_h^2$, and the pdfs for $n_f+1$ flavors are defined from the light
flavor pdfs and the massive operator matrix elements for $n_f$ light flavors.
In Mellin $n$-moment space, the heavy quark pdfs can then be expressed
in terms of the original light ones at NNLO as
\begin{equation}
(h+\bar{h})_{n_f+1}(n,m_h^2) = a_s^2
\Big[\tilde{A}_{hq}^{\rm PS,(2)}(n)\,\,
\Sigma_{n_f}(n,m_h^2)+\tilde{A}_{hg}^{\rm S,(2)}(n)\,\,
g_{n_f}(n,m_h^2)\Big],
\end{equation}
$(h-\bar{h})_{n_f+1}(n,m_h^2)=0$ and the remaining matching conditions
for the light pdfs and the gluon distribution read
\begin{equation}
(q\pm \bar{q})_{n_f+1} (n,m_h^2) = (q\pm \bar{q})_{n_f}(n,m_h^2)
 +a_s^2\,\, A_{qq,h}^{\rm NS,(2)}(n)(q\pm \bar{q})_{n_f}(n,m_h^2)
\end{equation}
\begin{equation}
g_{n_f+1}(n,m_h^2) = g_{n_f}(n,m_h^2)+a_s^2
\Big[A_{gq,h}^{\rm S,(2)}(n) \Sigma_{n_f}(n,m_h^2)
+ A_{gg,h}^{\rm S,(2)}(n) g_{n_f}(n,m_h^2)\Big]
\end{equation}
with the moments of the flavor singlet quark distribution being given
by
\begin{equation}
\Sigma_{n_f}(n,Q^2) = \int_0^1 dx\,\, x^{n-1}
\sum_{k=1}^{n_f}\,\, [q_k(x,Q^2)+\bar{q}_k(x,Q^2)]
\end{equation}
where $q_1\equiv u$, $q_2\equiv d$, etc. The coefficients $A^{(2)}(x)$
of the operator matrix elements have been originally calculated in 
\cite{ref16} and their Mellin moments $A^{(2)}(n)$ have been analyzed
and given in \cite{ref8,ref11,ref30}. 
Due to our choice $Q^2=m_h^2$ for the thresholds, only the scale-independent
parts of the expressions for $A^{(2}(n)$ are needed which, for completeness,
will be summarized in the Appendix.  
The strong coupling $a_s\equiv \alpha_s(Q^2)/4\pi$ is matched at the 
various thresholds $Q^2=m_h^2$ in the standard way as recapitulated in
\cite{ref1} with $\alpha_s(M_Z^2)=0.1124$ as obtained in our dynamical
scenario \cite{ref1} using $m_c=1.3$ GeV, $m_b=4.2$ GeV, and $m_t=175$
GeV.
Our choice for the input of the `heavy' VFNS distributions in (1) 
are the unique NNLO dynamical FFNS distributions \cite{ref1} at
$Q^2=m_c^2$, as obtained from the NNLO evolution of our valencelike
input distributions at $Q^2=\mu^2=0.55$ GeV$^2$ (see Table I of 
\cite{ref1}.
The resulting VFNS predictions at scales $Q^2\gg m_h^2$ should become
insensitive to this input selection \cite{ref14}, since asymptotically
the VFNS pdfs are dominated by their radiative evolution rather than 
by the specific input at $Q^2=m_h^2$, i.e., because of the long evolution
distance input differences get evolved away at $Q^2\gg m_h^2$ where the
universal perturbative QCD splittings dominate.

For illustration we show in Fig.\ 1 our NNLO charm and bottom 
distributions together with the fully convoluted $F_2^c$ and $F_2^b$
structure functions which are also compared with the NLO ones. 
In general the NNLO results for $F_2^{c,b}$ fall below the NLO ones
(dash--dotted curves) in the small-$x$ region.  
Here at NNLO the
${\cal{O}}(\alpha_s^2)$ convolutions of the fermionic and gluonic
coefficient functions with $\stackrel{(-)}{h}$ and the gluon distribution,
respectively, become more important than at NLO since the  `heavy' quark
distributions $xc$ and $xb$ by themselves (short-dashed curves) are
sizably different from $\frac{9}{8} F_2^c$ and $\frac{9}{2} F_2^b$, 
respectively.
At NLO the ${\cal{O}}(\alpha_s)$ the quark and gluon convolution 
contributions almost cancel \cite{ref12} and thus $xc$ and $xb$ almost
coincide with the appropriate NLO structure functions in Fig.\ 1.
As is obvious from Fig.\ 1, however, such differences between NNLO
and NLO results lie always within the $1\sigma-2\sigma$ uncertainty
bands in the relevant large $Q^2$ region, 
$Q^2$ \raisebox{-0.1cm}{$\stackrel{>}{\sim}$} $10^2-10^3$ GeV$^2$,
and can therefore be hardly delineated experimentally.

The shape of the gluon distribution at two typical fixed values of $x$,
relevant for Higgs boson production at LHC, is illustrated in Fig.\ 2.
At small to medium values of $Q^2$ the NNLO gluon falls always
{\em below} the NLO one and in both orders the gluon remains positive
at small $Q^2$ in the very small-$x$ region.  
This dampening of the NNLO gluon is a typical NNLO effect being mainly
caused \cite{ref1} by the gluonic 3-loop splitting function $P_{gg}^{(2)}$
which is {\em negative} and {\em more} singular $(\sim -\frac{1}{x}\ln 
\frac{1}{x})$ in the small-$x$ region \cite{ref31} than the NLO (and LO)
ones.
At large values of $Q^2$ the NNLO and NLO gluon distributions become
practically indistinguishable.

\subsection
{Weak gauge boson production}

As a next test of our VFNS distributions we turn to the hadronic $W^{\pm}$
and $Z^0$ production.  The inclusive differential cross section is
usually written as \cite{ref32}
\begin{equation}
\frac{d\sigma^V}{dQ^2} = \tau \sigma_V(Q^2,\, M_V^2)\, W_V(\tau,Q^2)\,\,,
\quad \tau = Q^2/s
\end{equation}
where V is one of the gauge bosons of the Standard Model ($\gamma,\, Z^0$
or $W^{\pm}$) which subsequently decays into a lepton pair ($\ell_1 \ell_2$)
with invariant mass $M_{\ell_1\ell_2}$, i.e. $Q^2\equiv M_{\ell_1\ell_2}^2$,
and $\sigma_V$ is the pointlike cross section, e.g., 
$\sigma_\gamma = 4\pi\alpha^2/9Q^4$,
etc. \cite{ref32}.  The hadronic Drell--Yan structure function is 
represented by
\begin{equation}
W_V(\tau,Q^2) = \sum_{i,j} \int_\tau^1 \frac{dx_1}{x_1}\, 
 \int_{\tau/x_1}^1 \frac{dx_2}{x_2}\, PD_{ij}^V(x_1,\, x_2,\, \mu_F^2)\,
  \Delta_{ij}(\frac{\tau}{x_1,x_2},\, Q^2,\, \mu_F^2)
\end{equation}
with $PD_{ij}^V$ denoting the usual combination of pdfs of flavor type
$i$ and $j$ which depend on the factorization scale $\mu_F$.  The QCD
correction term is expanded in a power series of $\alpha_s$ (or
$\alpha_s/4\pi$ or $\alpha_s/\pi)$  as follows
\begin{equation}
\Delta_{ij}(x,Q^2,\mu_F^2) = \sum_{n=0}^2 \alpha_s^n(\mu_R^2)\, 
 \Delta_{ij}^{(n)}(x,Q^2,\mu_F^2,\mu_R^2)
\end{equation}
with $\Delta_{ij}^{(1)}$ and the NNLO 2-loop $\Delta_{ij}^{(2)}$ being
given in \cite{ref32,ref33}, and the choice for the renormalization
scale $\mu_R=\mu_F$ is dictated by all presently available pdfs which
have been determined and evolved according to $\mu_R=\mu_F$. 
The scale uncertainties of our predictions are defined by taking 
$M_W/2\leq \mu_F\leq 2 M_W$, using $M_W=80.4$ GeV (and similarly for
$Z^0$ production, using $M_Z= 91.2$ GeV).  
Furthermore, it should be noted that only the initial $u,d,s,c$ quark
flavors and the gluon contribute sizably via the various fusion
subprocesses in (6) 
to the production rates of gauge bosons, whereas all subprocesses 
involving the $b$-flavor distribution, e.g., $u\bar{b}\to W^+$,
$\bar{c}b\to W^-$, etc., are negligibly small \cite{ref12}.

Our NNLO predictions for $\sigma(p\bar{p}\to W^{\pm}X)$ and 
$\sigma(p\bar{p}\to Z^0X)$ are compared with our NLO ones \cite{ref12}
in Fig.\ 3 where, for comparison, we also show the predictions of
Alekhin \cite{ref14,ref34}.  
The vector boson production rates at NNLO
are typically slightly {\em larger} (by more than $1\sigma$ than at
NLO with a $K\equiv$ NNLO/NLO factor of $K^{W^+ +W^-} = 1.04$ and
$K^{Z^0}=1.06$ at Tevatron energies ($\sqrt{s}=1.96$ TeV, cf.\ Table 1),
to be compared with the predictions of Alekhin \cite{ref14,ref34}
$K_A^{W^+ +W^-}\simeq K_A^{Z^0}=1.03$.  This confirms again the fast
perturbative convergence at NNLO since the NLO/LO $K$-factor \cite{ref12}
is 1.3 for $W^+ +W^-$ production at $\sqrt{s}=1.96$ TeV.
Our predicted NNLO cross sections at $\sqrt{s}=1.96$ TeV (cf. Table 1),
$\sigma(p\bar{p}\to W^+ +W^- +X)=25.2$ $nb$ and 
$\sigma(p\bar{p}\to Z^0+X)=7.5$ $nb$, are similar to the ones of
MSTW \cite{ref35}, 25.4 $nb$ and 7.4 $nb$, respectively, but smaller
than the ones obtained by Alekhin \cite{ref34}, 25.8 $nb$ and 7.8 $nb$,
respectively.
For the latter cases the branching ratios $B(W\to\ell\nu)=0.108$
and $B(Z\to\ell^+\ell^-)=0.034$ have been used. 
It is obvious from Fig.\ 3 that most of these results are within
the present experimental $1\sigma$ uncertainty. The scale uncertainties
of our NNLO predictions in Fig.\ 3 at $\sqrt{s} = 1.96$ TeV amount to
less than 0.5\% (where $\mu_F=M_V/2$ gives rise to the upper limits
and $\mu_F=2M_V$ to the lower limits, with $V=W^{\pm},Z^0)$ which is
four times less than at NLO \cite{ref12}.

In Table 2 we present our NNLO predictions for $W^{\pm}$ and $Z^0$ 
production at LHC energies.  
For comparison we also display our previous NLO results
\cite{ref12}. Here the scale uncertainties amount to less than 1.7\%,
i.e., are about half as large than the stated pdf uncertainties and 
than the scale uncertainties at NLO \cite{ref12}.  For example, the
full NNLO expectations at $\sqrt{s}=14$ TeV are
\begin{eqnarray}
\sigma(pp\to W^+ +W^- +X) & = & 190.2 
         \pm 5.6_{\rm pdf}\,\,\,\,^{+1.6}_{-1.2}|_{\rm scale}\,\,nb
\\
\sigma(pp\to Z^0 +X) & = & \,\,\,55.7 \pm
             1.5_{\rm pdf}\,\,\,\,^{+0.6}_{-0.3}|_{\rm scale}\,\,nb\, .
\end{eqnarray}
Here the scale choice $\mu_F=2M_V$ gives rise to the upper limits, and
$\mu_F=M_V/2$ to the lower limits of our predicted cross sections.
These results are about 5\% smaller than the ones of MSTW \cite{ref35},
whereas Alekhin \cite{ref34} obtained 195.2 $nb$ and 57.7 $nb$ for 
$W^+ +W^-$ and $Z^0$ production, respectively, with similar pdf 
uncertainties as in (8) and (9).
>From Table~2 it becomes obvious that the vector boson production rates
somewhat increase at NNLO as compared to the NLO expectations, but such
differences are well within present pdf and scale uncertainties. 
Moreover, the smallness of such differences ($K\simeq 1.02$) indicates
the reliability of perturbative predictions already at NLO. For
comparison we note that within the FFNS (where the heavy $c,b,t$
quark flavors do not form massless partons of the nucleon) the 
$W^+ +W^-$ production rate has been estimated \cite{ref12} to be about
192.7 $nb$ at NLO with a total (pdf as well as scale) uncertainty of
about 5\%.  
In general the NLO--VFNS prediction of 186.5 $nb$ in Table~2 falls
somewhat below that estimate but remains well within its total
uncertainty of about 6\% \cite{ref12}.  Due to the reduced scale
ambiguity at NNLO and due to the slightly different NNLO estimates
obtained by other groups as discussed above, we conclude that the rates
for gauge boson production at LHC energies can be rather confidently
predicted with an accuracy of about 5\% irrespective of the factorization
scheme.

\subsection{Higgs boson production}

As a final application of our NNLO VFNS pdfs we consider the hadronic
production of the SM Higgs boson.  Similar as for gauge boson production
in Sec.\ 2.1, the total inclusive cross section for Higgs boson
production is usually written as \cite{ref32,ref33,ref40,ref41}
\begin{equation}
\sigma^H(s)=\sum_{i,j} \int_0^1 dx_1\, dx_2\, PD_{ij}^H\, 
    (x_1,\, x_2,\, \mu_F^2)\, 
  \hat{\sigma}_{ij\to H}\, (\hat{s}=x_1 \, x_2 \, s,\, \mu_F^2,\ 
    \mu_R^2)
\end{equation}
with the partonic cross sections for $ij\to HX$ being written as
\begin{equation}
\hat{\sigma}_{ij\to H}(\hat{s}) = \sigma_0\Delta_{ij}(\hat{s}) =
 \sigma_0 \sum_{n=0}^2\,  \alpha_s^n\, (\mu_R^2)\, \Delta_{ij}^{(n)}
  (\hat{s},\,  \mu_F^2,\,  \mu_R^2)
\end{equation}
where the obvious $\mu_F$ and $\mu_R$ dependencies have been suppressed.
The dominant Higgs production proceeds via gluon-gluon fusion where
$\sigma_0 = \alpha_s^2/(576\, \pi v^2)$ for the initial LO process
$gg\to H$, with the Higgs vev $v=(\sqrt{2}\, G_F)^{-1/2}\simeq 246$ GeV,
and the NLO QCD corrections $\Delta_{ij}^{(1)}$ are given in 
\cite{ref42,ref43} and the NNLO $\Delta_{ij}^{(2)}$ ones in 
\cite{ref33,ref40}.  
The factorization scale is ususally chosen to be $\mu_F=M_H$, and 
the scale uncertainty is illustrated by taking $\frac{1}{2} M_H
\leq \mu_F \leq 2 M_H$.
The much smaller contribution stemming from bottom quark annihilation
starts at LO with $b\bar{b}\to H$ where $\sigma_0=\pi\lambda_b^2/
(12 M_H^2)$ with $\lambda_b=\sqrt{2} m_b/v$ in the SM, and the NLO and
NNLO QCD corrections $\Delta_{ij}^{(1)}$ and $\Delta_{ij}^{(2)}$ can
be found in \cite{ref41}. 
Here it has been argued \cite{ref44,ref45,ref46} that the optimal 
choice of the factorization scale is $\mu_F\simeq M_H/4$ where the
differences between the VFNS and the FFNS are significantly reduced
and the LO, NLO and NNLO results become rather similar \cite{ref41},
which implies a more stable perturbative behavior. 
The scale uncertainty is again probed by taking $\frac{1}{2}(M_H/4)
\leq \mu_F\leq 2 (M_H/4)$. 
In both cases $\mu_R=\mu_F$ as dictated by all presently available
pdfs.

In Fig.\ 4 we show, as a function of the Higgs mass, our NNLO (thick
solid curve) and NLO (thick dash-dotted curve) predictions
for LHC for Higgs boson production via the dominant gluon-gluon fusion
subprocess which starts, at LO, with $gg\to H$.  The shaded regions
around these central predictions are due to the $\pm 1\sigma$ pdf 
uncertainties.  
Reducing the scale to $\frac{1}{2} M_H$ one arrives at the thin upper
curves at each order, whereas the scale choice $\mu_F=2M_H$ results in 
the respective lower curves, where the appropriate $\pm 1\sigma$ pdf
ambiguities have also been included for each choice of scale. 
These ambiguities for each scale choice $\mu_F=\mu_R=\frac{1}{2}M_H$,
$M_H$, $2M_H$ are more explicitly illustrated in Table 3. 
Despite the fact that the NLO and NNLO total uncertainty bands overlap
in Fig.\ 4, the predicted NNLO production rates are typically about
20\% {\em larger} than at NLO.
The insensitivity of these predictions with respect to the appropriate 
choice of the pdfs is illustrated by the dashed curve which has been
obtained by using NNLO matrix elements and (inconsistently) NLO pdfs.
Here for the dominant gluon fusion process such an inconsistent choice
of the pdfs appears to be immaterial and the production rates depend
dominantly on the NNLO QCD dynamics.  Our central predictions in Fig.\ 4
are comparable with the ones presented in \cite{ref47}, but are about
10\% smaller than the ones in \cite{ref35}.  
For completeness we also show in Fig.\ 5 our NNLO expectations for
Higgs boson production at the Tevatron, $\sqrt{s}=1.96$ TeV. 
Note that here the total uncertainty bands almost double at NNLO and 
NLO as compared to the ones at LHC in Fig.\ 4.

In Fig.\ 6 we finally show the subdominant contribution to Higgs boson
production at LHC due to bottom-quark fusion which starts with 
$b\bar{b}\to H$ at LO.  
Here, in contrast to the by far dominant gluon fusion process in Fig.ß 4,
the NNLO and NLO predictions, together with their $\pm 1\sigma$ pdf
uncertainties, almost coincide with the NNLO results falling very 
slightly below the NLO ones. 
Here, however, the correct choice of the NNLO pdfs turns out to be
important, since choosing (incorrectly) NLO pdfs \cite{ref12} for a
NNLO analysis results in too small a production rate as shown by the
dashed curve.  
At NNLO the scale dependence is here, again in contrast to the by far
dominant gluon-gluon fusion process, very marginal: using $\mu_F=\mu_R
=2(M_H/4)$ instead of $M_H/4$ leaves the results in Fig.\ 6 practically
unchanged, whereas the choice $\mu_F=\mu_R=\frac{1}{2}(M_H/4)$
increases the results by at most 5\%.

It should be again emphasized that here, as in the previous case of 
gauge boson production, the simpler VFNS yields sufficiently reliable
predictions for Higgs boson production despite the fact that a fully
massive FFNS analysis cannot be performed at NNLO at present (due to
the absence of NNLO, and in many cases even NLO, matrix elements with
$m_h\neq 0$).
This is due to the fact that $\sqrt{\hat{s}_{th}}/m_b=(2m_b + M_H)/
m_b\gg 1$, i.e., nonrelativistic contributions from the threshold
region in the FFNS are suppressed, and thus the FFNS and VFNS predictions
should not differ too much \cite{ref12}, as has been discussed in more
detail in the Introduction.  
Indeed it has been noted \cite{ref12} that the FFNS and VFNS results
at NLO are compatible \cite{ref48,ref49,ref50}, and that the VFNS rates
exceed the corresponding FFNS Higgs boson production rates by about
10-20\%, depending on the choice of the scale $\mu_F=\mu_R$.

\section{Summary and Conclusions}

Based on our recent NNLO dynamical parton distributions as obtained in
the FFNS \cite{ref1}, we generated radiatively VFNS parton distributions
at NNLO where also the heavy quark flavors ($c,b,t$) become massless 
partons within the nucleon.  
The latter pdfs in the  `variable flavor number' factorization scheme
considerably ease the otherwise unduly complicated calculations in the
FFNS where for the time being fully massive NNLO analyses are not 
possible (and in many cases even not at NLO), such as the calculation
of gauge- and Higgs-boson production and heavy quark production at
collider energies.
It has been shown \cite{ref12} that for situations where the invariant
mass of the produced system exceeds by far the mass of the participating
heavy flavor in the FFNS, the VFNS predictions deviate rather little
from the FFNS ones, typically by about 10\% which is within the margins
of renormalization and factorization scale uncertainties and ambiguities
related to presently available parton distributions.  
As an application of our NNLO VFNS pdfs we studied the perturbative
stability of the predictions for gauge ($W^{\pm},\, Z^0$) and SM Higgs
boson production at collider energies by comparing them with the 
appropriate NLO results, taking into account pdf uncertainties as 
well as scale dependencies.  
The NNLO predictions for gauge boson production are typically slightly
{\em larger} (by more than  $1\sigma$) than the NLO ones, cf.\ Table 1
and 2. 
Due to the reduced scale ambiguity at NNLO and due to the slightly
different NNLO estimates of other groups we conclude that the rates
for gauge boson production at LHC energies can be rather confidently
predicted with an accuracy of about 5\%.
At the Tevatron ($\sqrt{s}=1.96$ TeV) the NNLO predictions are more
than a $2\sigma$ pdf uncertainty above the NLO ones, but most of these
results are within the present experimental $1\sigma$ uncertainty.

The NNLO predictions for the production of the SM Higgs boson via
the dominant gluon fusion (in contrast to the subdominant bottom-quark
fusion) process are, at collider energies, typically about 
20\% {\em larger} than at
NLO, but their respective total (pdf and scale) uncertainty bands
overlap.
Higgs boson production at LHC ($\sqrt{s}=14$ TeV) can be predicted with
an accuracy of about 10\% at NNLO (with the total uncertainty being 
almost twice as large at NLO), whereas the uncertainty almost doubles
at Tevatron ($\sqrt{s}=1.96$ TeV).
\vspace{0.25cm}

A FORTRAN code (grid) containing our NNLO-VFNS pdfs (including their
uncertainties) can be obtained on request or directly online from
http://doom.physik.uni-dortmund.de/pdfserver.
\vspace{0.75cm}

\noindent{\large{\bf Acknowledgements}}\\
We are grateful to R.~Harlander for providing us with the NNLO routine
for calculating hadronic Higgs boson production cross sections, as well
as for a clarifying correspondence.  
We also thank J.~Bl\"umlein and M.~Gl\"uck for helpful discussions and 
comments.
This work has been supported in part by 
the  `Bundesministerium f\"ur Bildung und Forschung', Berlin/Bonn.
\vspace{1.25cm}

\noindent{\large{\bf Note added.}}\\
While completing this manuscript, an investigation along similar lines
appeared 
[S.~Alekhin, J.~Bl\"umlein, S.~Klein, and S.~Moch, arXiv:0908.2766].
Several results are similar to ours.  However, the gauge boson production
rates are about 4\% larger at the Tevatron and about 10\% larger at LHC
than our NNLO predictions.  
Similarly, the NNLO predictions for Higgs boson production at LHC
($\sqrt{s}=14$ TeV) are 5--8\% larger for 
$M_H$ \raisebox{-0.1cm}{$\stackrel{<}{\sim}$} 150 GeV than ours, but
agree with us for larger Higgs masses, whereas at the Tevatron their
expected rates are 12-30\% smaller than our ones for $M_H=$100--200 GeV.
The comparison of these production rates refers always to the central
results, disregarding all pdf and scale uncertainties.

\newpage
\appendix
\section*{Appendix}
\renewcommand{\theequation}{A.\arabic{equation}}
\setcounter{equation}{0}
\noindent Some analytic form of the Mellin $n$-moments of the operator matrix 
elements including the heavy quark flavors, which have been originally
calculated in Bjorken-$x$ space \cite{ref16} and which are needed in
(1)--(3), 
have been already implicitly used in some NNLO evolution programs
(see, e.g.\ \cite{ref51}).
Here we summarize the relevant analytic expressions which can be directly
continued to complex values of $n$ as required for the Mellin inversions
to Bjorken-$x$ space.

The Bjorken-$x$ expressions are given in Appendix B of \cite{ref16}
and we follow the notation used there [cf.\ also (1)--(3)]. 
Due to our choice $Q^2=m_h^2$ for the flavor transition thresholds, only
the scale-independent parts of these expressions contribute. 
Their moments are as follows:
\begin{eqnarray}
\frac{1}{C_F T_f}\tilde{A}^{PS,(2)}_{hq}(n) & = & 
 -8\, \frac{n^4+2n^3+5n^2+4n+4}{(n-1)n^2(n+1)^2(n+2)} S_2(n-1)
  -\frac{448}{27} \frac{1}{n-1} -\frac{44}{n} +\frac{48}{n^2}
      -\frac{4}{n^3}\nonumber
\\
& & +\, \frac{24}{n^4}
    -\frac{12}{n+1}+\frac{56}{(n+1)^2}+\frac{28}{(n+1)^3}
     +\frac{24}{(n+1)^4}+\frac{1960}{27}\, \frac{1}{n+2}\nonumber 
\\
& & +\, \frac{448}{9}\, \frac{1}{(n+2)^2}  +\frac{64}{3}\, 
       \frac{1}{(n+2)^3}
\end{eqnarray}
where $C_F=\frac{4}{3}$, $T_f=\frac{1}{2}$ und 
$S_k(n)\equiv\sum_{j=1}^n\,\frac{1}{j^k}$ using 
\begin{equation}
S_1(n)=\psi(n+1)+\gamma_E,\quad
 S_{k'}(n) = \frac{(-1)^{k'-1}}{(k'-1)!} \, \psi^{(k'-1)}(n+1)+\zeta(k'),
 \,\,\, k'\geq 2,
\end{equation}
with $\psi^{(i)}(z) = d^{(i+1)}\ln\Gamma(z)/dz^{i+1}$ and
$\gamma_E=0.5772156649$, $\zeta(2)=\pi^2/6$ and 
$\zeta(3) = 1.2020569032$ for the analytic contiunation to complex $n$.
Since the moment of the  rather complicated coefficient 
$\tilde{A}_{hg}^{S,(2)}$ appearing in (1), and
as given in (B.3) of \cite{ref16}, cannot be straightforwardly expressed
in terms of analytic functions of $n$ \cite{ref8}, we have employed for
practical purposes the $n$-moment of the sufficiently accurate 
$x$-parametrization suggested in \cite{ref51}:
\begin{eqnarray}
\tilde{A}_{hg}^{S,(2)}(n) & = &   
 1.111 \frac{S_1^3(n)}{n} -0.4\frac{S_1^2(n)}{n}
   +(\frac{2.77}{n}+\frac{293.6}{n^3})S_1(n)+\frac{3.333}{n}S_1(n)S_2(n)
\nonumber
\\
& & -\big(\frac{0.4}{n}-\frac{293.6}{n^2}\big)\,
     S_2(n)+295.822\, \frac{S_3(n)}{n}
      -0.006 - \frac{24.89}{n-1}\nonumber
\\
& & -\frac{187.8+293.6\zeta(3)}{n}+\frac{93.68-293.6\zeta(2)}{n^2}
      -\frac{6.584}{n^3}\!+\!\frac{9.336}{n^4}\!+\!\frac{249.6}{n+1}\,  .
\end{eqnarray}
The remaining coefficients relevant for the light quark and gluon sector
in (2) and (3) 
respectively, can be straightforwardly transformed to $n$-space :
\begin{eqnarray}
\frac{1}{C_F T_f}\, A_{qq,h}^{\rm NS,(2)}(n) & = &
 -\frac{224}{27}\, S_1(n-1)+\frac{40}{9}\ S_2(n-1)-\frac{8}{3}\, S_3(n-1)
   +\frac{73}{18} +\frac{44}{27}\, \frac{1}{n}-\frac{4}{9}\,
      \frac{1}{n^2}\nonumber
\\
& & -\frac{268}{27}\, \frac{1}{n+1} + \frac{44}{9}\, \frac{1}{(n+1)^2}
    -\frac{4}{3}\, \frac{1}{n^3} -\frac{4}{3}\, \frac{1}{(n+1)^3}
\end{eqnarray}
\begin{eqnarray}
\frac{1}{C_F T_f}\, A_{gq,h}^{S,(2)}(n) & = &
     \frac{8}{3}\, \frac{1}{n-1}
      \Big[ S_1^2(n-1)-\frac{10}{3}\, S_1(n-1)+S_2(n-1)+\frac{56}{9}\Big]\nonumber
\\
& & 
   -\frac{8}{3}\, \frac{1}{n}
    \Big[S_1^2(n)-\frac{10}{3}\, S_1(n)+S_2(n)+\frac{56}{9}\Big]\nonumber
\\
& & +\frac{4}{3}\, \frac{1}{n+1}
   \Big[S_1^2(n+1)-\frac{16}{3}\, S_1(n+1)+S_2(n+1)+\frac{86}{9}\Big]
\end{eqnarray}
\begin{eqnarray}
A_{gg,h}^{S,(2)}(n) &  = &  
    4C_F T_f\Big[ -\frac{15}{4}\, -\, \frac{2}{n-1}\,  +\, \frac{20}{n}\, -\, \frac{8}{n^2}\, 
       +\, \frac{3}{n^3}\, -\, \frac{2}{n^4}\, -\, \frac{12}{n+1}\, -\, \frac{12}{(n+1)^2}\, 
        \nonumber
\\
& & \quad\quad\quad\quad   +\, \frac{5}{(n+1)^3}\, -\frac{2}{(n+1)^4}\, - \frac{6}{n+2}\Big]\nonumber
\\
& & + 4C_A T_f\Big[-\frac{56}{27}\, S_1(n-1)+\, \frac{1}{3}\, \frac{S_1(n+1)}{n+1}\, 
     +\, \frac{5}{18}\, + \frac{139}{27}\, \frac{1}{n-1}\, - \,\frac{157}{27n}\, 
           - \frac{13}{9n^2}\nonumber
\\
& &  \quad\quad\quad\quad  +\, \frac{2}{3n^3}\, 
       +\,  \frac{137}{27} \frac{1}{n+1}- \frac{22}{9}\frac{1}{(n+1)^2}
        +\frac{2}{3} \frac{1}{(n+1)^3} - \frac{175}{27} \frac{1}{n+2}\Big]
\end{eqnarray}
with $C_A=3$.

\newpage

\begin{table}
\begin{center}
\renewcommand{\arraystretch}{1.5}
\begin{tabular}{|c|c|c|}
\hline
\multicolumn{3}{|c|}{$\sigma^{\textrm{p}\bar{\textrm{p}}\rightarrow\textrm{V}X}$ (nb), $\sqrt{\textrm{s}}$=1.96 TeV}\\
\hline
V  &  NNLO  &  NLO  \\
\hline
$\textrm{W}^\pm$              & $12.6 \pm 0.1$ & $12.1 \pm 0.1$ \\
$\textrm{W}^+ + \textrm{W}^-$ & $25.2 \pm 0.3$ & $24.2 \pm 0.3$ \\
$\textrm{Z}^0$                & $ 7.5 \pm 0.1$ & $ 7.1 \pm 0.1$ \\
\hline
\end{tabular}
\caption{NNLO predictions for vector boson production at the Tevatron,
with the NLO ones being taken from \cite{ref12}. The errors refer to 
the $\pm 1\sigma$ uncertainties implied by our dynamical NNLO \cite{ref1}
and NLO \cite{ref7} pdfs. The scale uncertainties of our NNLO predictions,
due to $\frac{1}{2}M_V\leq\mu_F\leq 2M_V$, amount to less than 0.5\% 
(i.e., are about half as large as the stated pdf uncertainties) which is
about four times smaller than at NLO \cite{ref12}.}
\end{center}
\end{table}

\clearpage

\begin{table}
\begin{center}
\renewcommand{\arraystretch}{1.5}
\begin{tabular}{|c|c|c|}
\hline
\multicolumn{3}{|c|}{$\sigma^{\textrm{pp}\rightarrow\textrm{V}X}$ (nb), $\sqrt{\textrm{s}}$=10 TeV}\\
\hline
V  &  NNLO  &  NLO  \\
\hline
$\textrm{W}^+$               & $ 78.7 \pm 2.0$  & $ 76.7 \pm 1.7$ \\
$\textrm{W}^-$               & $ 55.8 \pm 1.4$  & $ 54.7 \pm 1.2$ \\
$\textrm{W}^+ + \textrm{W}^-$& $134.5 \pm 3.4$  & $131.6 \pm 2.9$ \\
$\textrm{Z}^0$               & $ 39.1 \pm 0.9$  & $ 38.1 \pm 0.8$ \\
\hline
\end{tabular}
\\[5mm]
\begin{tabular}{|c|c|c|}
\hline
\multicolumn{3}{|c|}{$\sigma^{\textrm{pp}\rightarrow\textrm{V}X}$ (nb), $\sqrt{\textrm{s}}$=14 TeV}\\
\hline
 V  &  NNLO  &  NLO  \\
\hline
$\textrm{W}^+$               & $109.8 \pm 3.2$  & $107.5 \pm 2.9$ \\
$\textrm{W}^-$               & $ 80.4 \pm 2.4$  & $ 79.1 \pm 2.1$ \\
$\textrm{W}^+ + \textrm{W}^-$& $190.2 \pm 5.6$  & $186.5 \pm 4.9$ \\
$\textrm{Z}^0$               & $ 55.7 \pm 1.5$  & $ 54.6 \pm 1.3$ \\
\hline
\end{tabular}
\caption{As in Table 1 but for LHC energies. The scale uncertainties of our
NNLO predictions amount to less than 1.7\% of the total predicted rates 
which is about half as large as the stated pdf $1\sigma$ uncertainties
and the scale uncertainties at NLO \cite{ref12}.}
\end{center}
\end{table}

\clearpage

\begin{table}
\begin{center}
\renewcommand{\arraystretch}{1.5}
\begin{tabular}{c|c|c|c|}
\begin{picture}(10,10)(1,-4)
\put(-7,-8){$M_H$}
\put(0,10){\line(1,-1){21}}
\put(7.1,5.2){$\mu_F$}
\end{picture}
      &    $\tfrac{1}{2}M_H$ &         $M_H$        &         $2M_H$      \\
\hline
$100$ & $67.6  \pm  3.0$ & $62.2  \pm  2.6$ & $57.3  \pm  2.2$\\
$150$ & $33.0  \pm  1.2$ & $30.4  \pm  1.0$ & $28.1  \pm  0.9$\\
$200$ & $19.7  \pm  0.6$ & $18.3  \pm  0.5$ & $16.9  \pm  0.5$\\
$250$ & $13.6  \pm  0.4$ & $12.6  \pm  0.4$ & $11.7  \pm  0.3$\\
$300$ & $10.7  \pm  0.3$ & $ 9.9  \pm  0.3$ & $ 9.2  \pm  0.3$\\
\hline
\end{tabular}
\caption{Typical NNLO scale dependencies of the cross sections
(in units of pb) for Higgs boson producion at $\sqrt{s}=14$ TeV via
the dominant gluon-gluon fusion subprocess with $M_H$ in GeV units.
The errors refer to the $1\sigma$ pdf uncertainties.  The maximal 
upper limits at $\mu_F=\mu_R=\frac{1}{2}M_H$ agree with the thin 
solid curve at NNLO in Fig.~4, whereas the lower curve in Fig.~4
corresponds to the minimal lower limits at $\mu_F=\mu_R=2M_H$.}
\end{center}
\end{table}
\clearpage
\begin{figure}
\begin{center}
\ifpdf
\includegraphics[width=14.0cm]{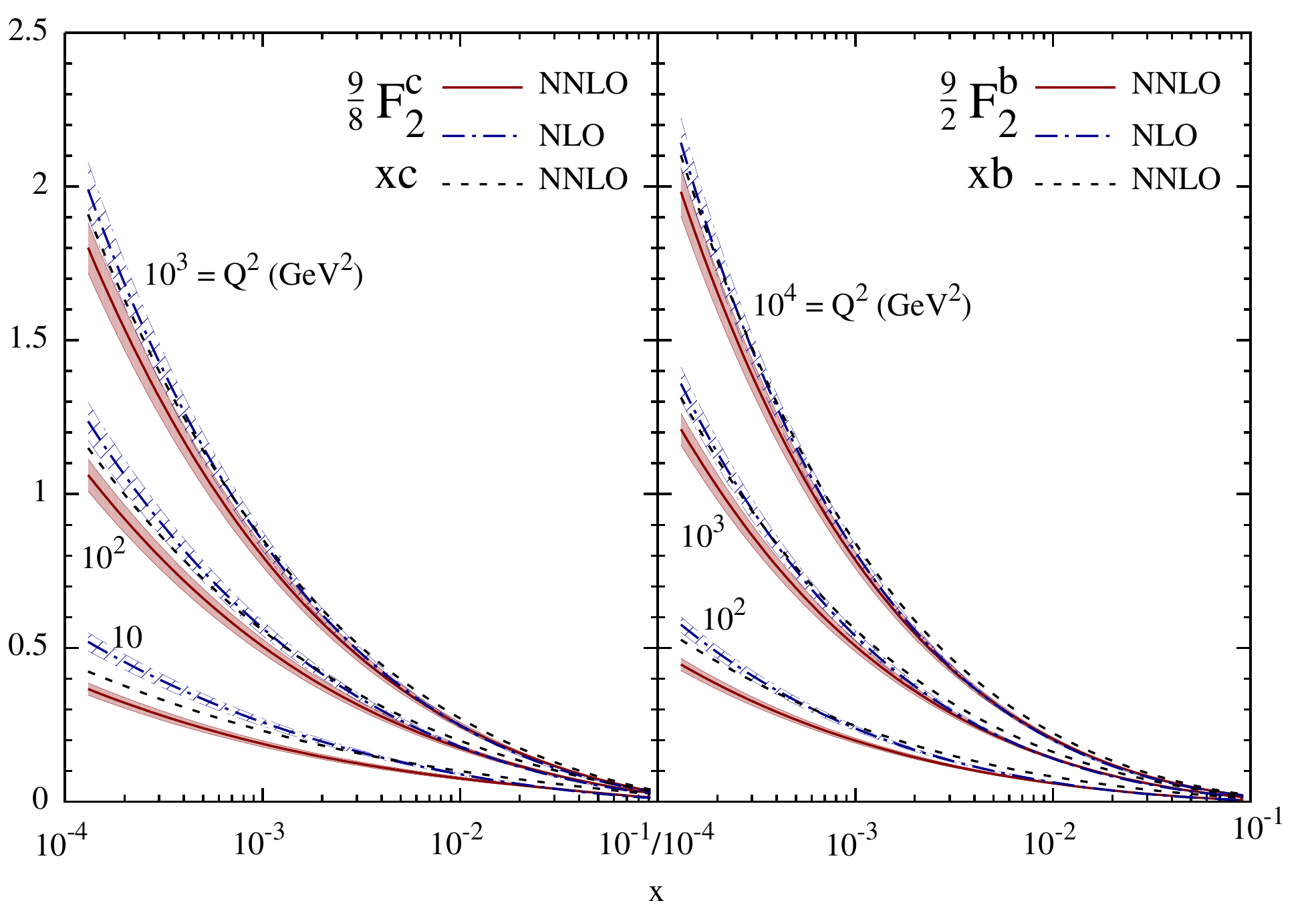}
\fi
\caption{The predicted $x$-dependencies of the charm and bottom quark
structure functions $\frac{9}{8}F_2^c(x,Q^2)$ and $\frac{9}{2}F_2^b(x,Q^2)$,
respectively, in the zero-mass VFNS, together with their $\pm 1\sigma$
uncertainties, at some typical fixed values of $Q^2$.  The NNLO charm
and bottom distributions, $xc(x,Q^2)$ and $xb(x,Q^2)$, are shown by the
short-dashed curves.}
\end{center}
\end{figure}
\clearpage
\begin{figure}
\begin{center}
\ifpdf
\includegraphics[width=14.0cm]{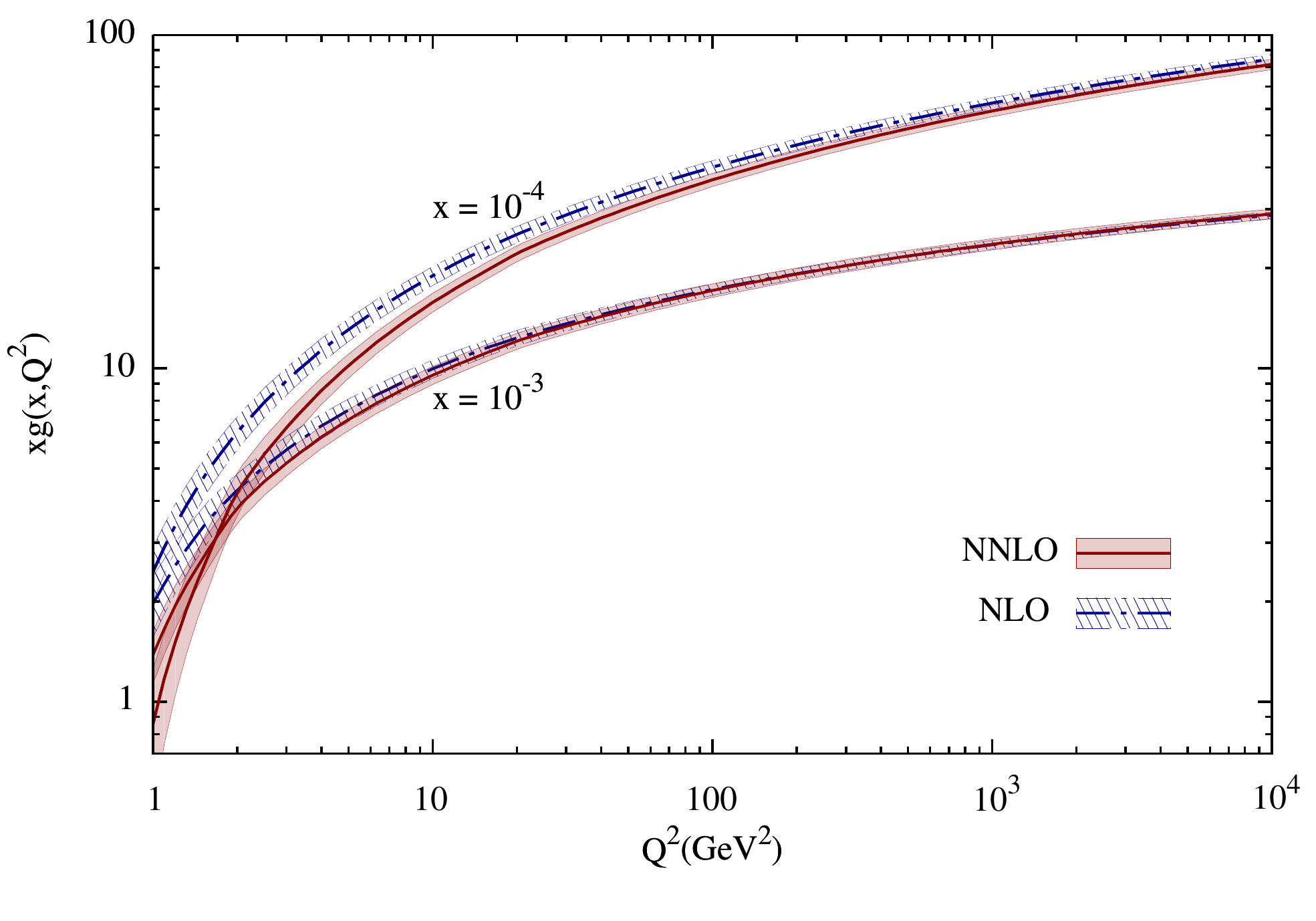}
\fi
\caption{The NNLO and NLO gluon distributions together with their 
$\pm 1\sigma$ uncertainty bands at two representative fixed values of $x$.} 
\end{center}
\end{figure}
\clearpage
\begin{figure}
\begin{center}
\ifpdf
\includegraphics[width=14.0cm]{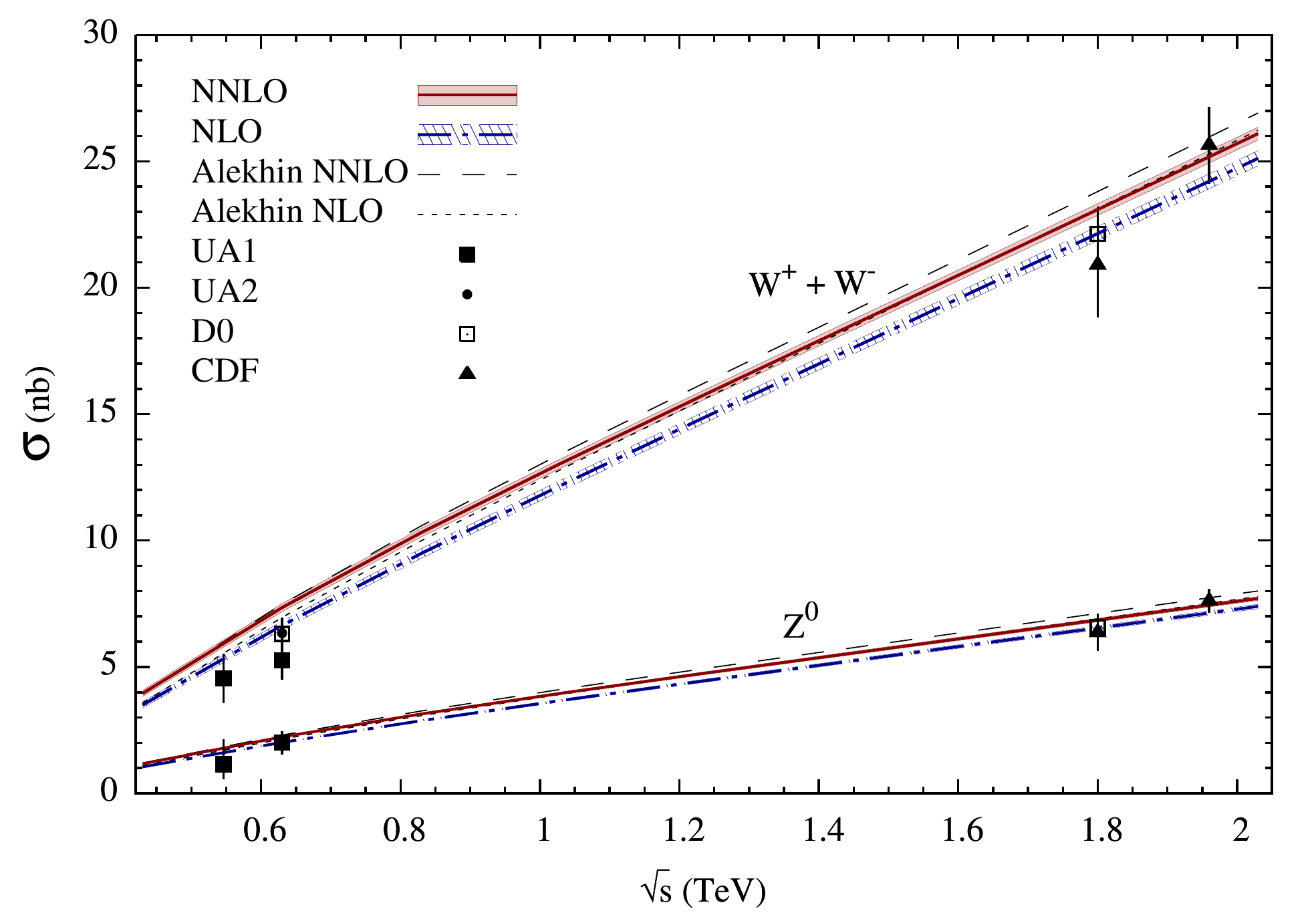}
\fi
\caption{Predictions for the total $W^++W^-$ and $Z^0$ production rates at
$p\bar{p}$ colliders with the data taken from \cite{ref36,ref37,ref38,ref39}.
Our NLO VFNS predictions are taken from \cite{ref12}, and the NLO and NNLO
ones of Alekhin from \cite{ref14,ref34}. The adopted momentum scale is
$\mu_F=\mu_R=M_V$ for $V=W^{\pm},Z^0$. The scale uncertainties of our NNLO
predictions, due to $\frac{1}{2} M_V\leq \mu_F\leq 2M_V$, amount to less
than 0.5\% at $\sqrt{s}=1.96$ TeV, i.e., is four times less than at NLO
\cite{ref12}.  The shaded band around our NNLO and NLO predictions are due
to the $\pm 1\sigma$ uncertainty implied by our dynamical NNLO \cite{ref1}
and NLO \cite{ref7} parton distributions.} 
\end{center}
\end{figure}
\clearpage
\begin{figure}
\begin{center}
\ifpdf
\includegraphics[width=14.0cm]{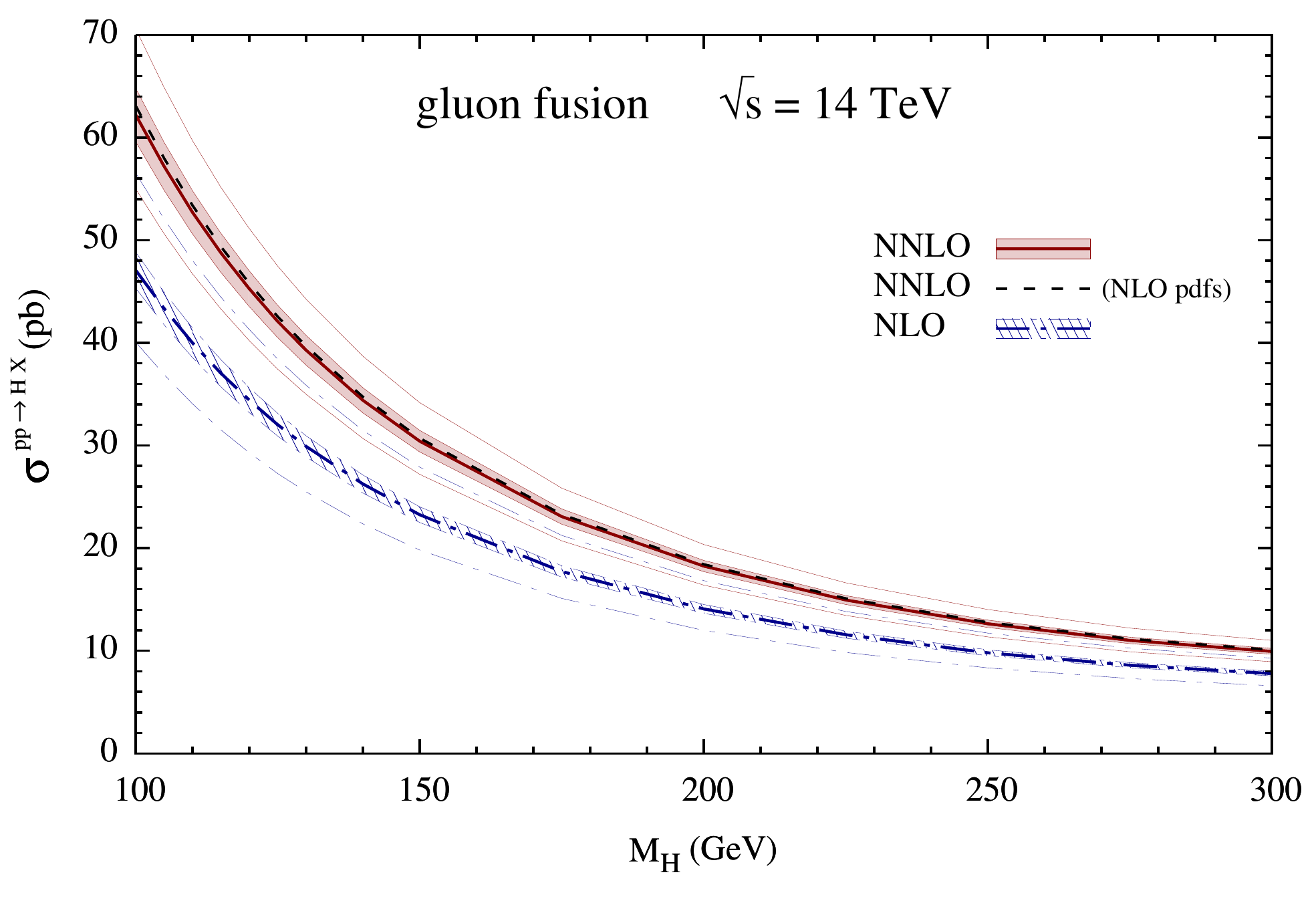}
\fi
\caption{Predictions for SM Higgs boson production at LHC ($pp\to HX$)
via the dominant gluon-gluon fusion process, which starts at LO with
$gg\to H$ via a top-quark loop.  The shaded bands around the central NNLO
and NLO predictions are due to the $\pm 1\sigma$ pdf uncertainties, all
referring to a scale choice $\mu_F=\mu_R=M_H$.  The thin solid and 
dash-dotted curves above these NNLO and NLO bands refer to a scale
$\mu_F=\mu_R=\frac{1}{2}M_H$ with $\pm1\sigma$ pdf uncertainties
included, and similarly the lower curves refer to $\mu_F=\mu_R=2 M_H$
(for more details cf.\ Table 3). The dashed NNLO curve is obtained by
using NNLO matrix elements and (inconsistently) NLO pdfs \cite{ref12}
with $\mu_F=\mu_R=M_H$.}
\end{center}
\end{figure}
\clearpage
\begin{figure}
\begin{center}
\ifpdf
\includegraphics[width=16.0cm]{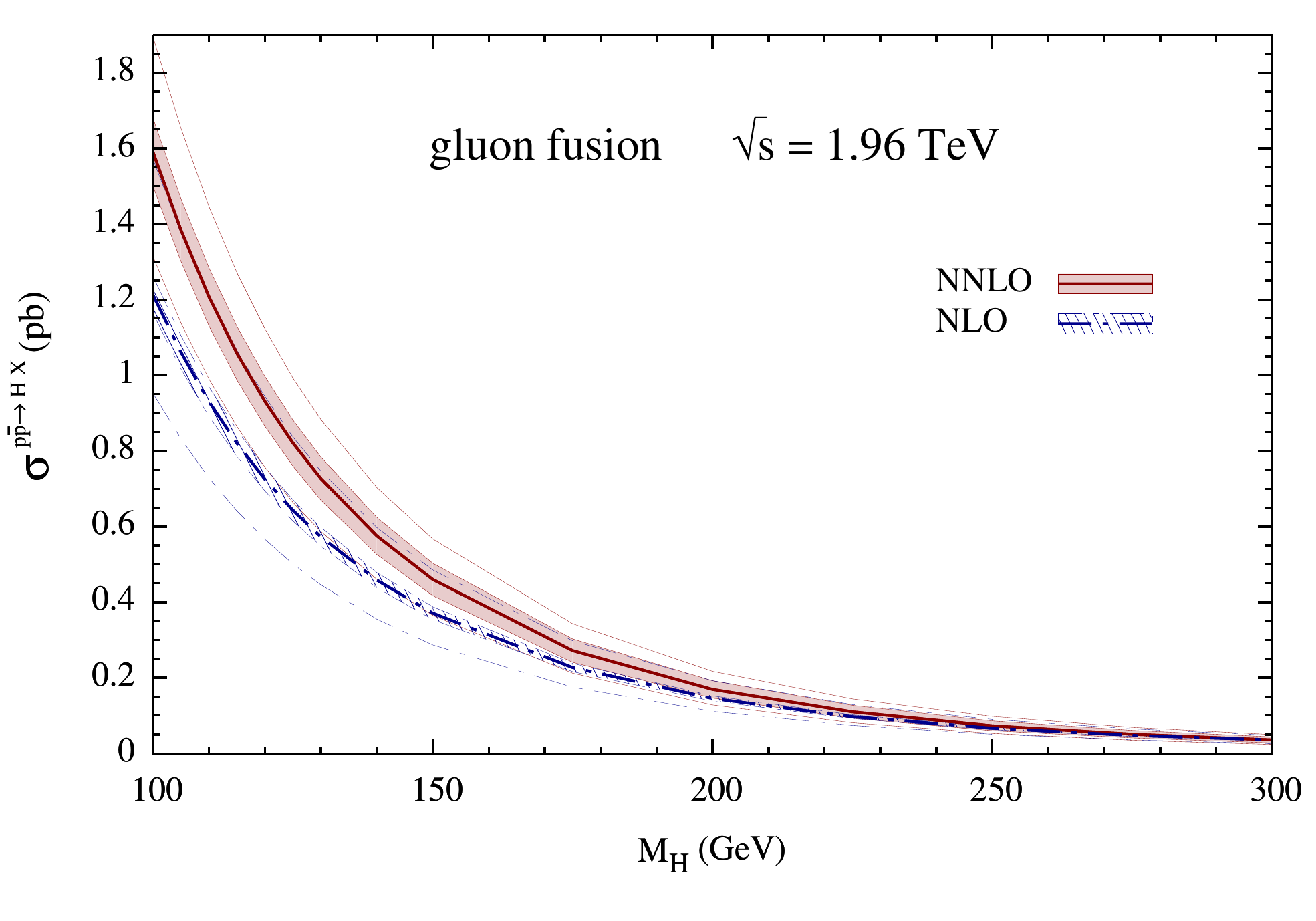}
\fi
\caption{As in Fig.\ 4 but for the Tevatron ($p\bar{p}\to HX$).}
\end{center}
\end{figure}
\clearpage
\begin{figure}
\begin{center}
\ifpdf
\includegraphics[width=16.0cm]{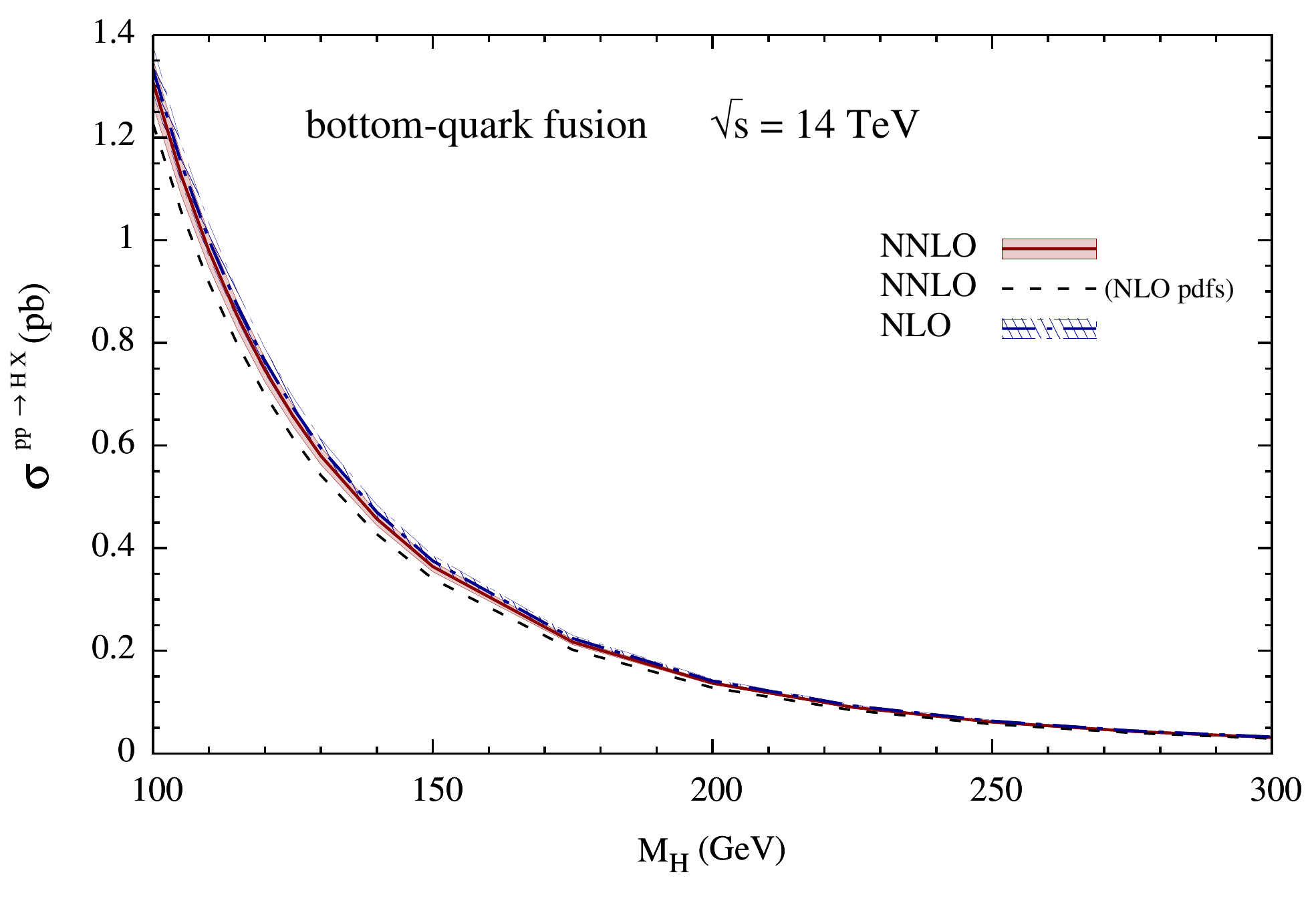}
\fi
\caption{Predictions for SM Higgs boson production at LHC via the 
small subdominant bottom-quark fusion process which starts with 
$b\bar{b}\to H$ at LO.  The shaded bands correspond to the $\pm 1\sigma$
pdf uncertainties of the NNLO and NLO central predictions, all referring
to a scale choice $\mu_F=\mu_R=M_H/4$. The dashed NNLO curve is obtained
by using NNLO matrix elements and (inconsistently) NLO pdfs \cite{ref12}.}
\end{center}
\end{figure}
\clearpage
\end{document}